\documentstyle[11pt,newpasp,twoside,psfig]{article}
\index{Chandra}
\index{supernova remnants!SN~1987A}
\index{supernovae!SN~1987A}
\index{supernova remnants!Magellanic Clouds}
\index{supernova remnants!morphology}
\index{supernova remnants!spectroscopy}

\def\cxo{{\em Chandra}}

\def\edcomment#1{\iffalse\marginpar{\raggedright\sl#1\/}\else\relax\fi}
\reversemarginpar

\begin{document}
\title{Monitoring The Evolution of SNR~1987A With \cxo}
 \author{Sangwook Park, David N. Burrows, Gordon P. Garmire}
\affil{Department of Astronomy and Astrophysics, 525 Davey Lab,
Pennsylvania State University, University Park, PA. 16802, USA}
\author{Svetozar A. Zhekov}
\affil{Space Research Institute, Moskovska str. 6, Sofia-1000, Bulgaria}
\author{Dick McCray}
\affil{JILA, University of Colorado, Box 440, Boulder, CO. 80309 USA}

\begin{abstract}
We report on the results of our monitoring program of SNR 1987A with 
the {\it Chandra X-Ray Observatory}. The high resolution images and 
the spectra from the latest {\it Chandra} data suggest that the 
blast wave has reached the dense inner ring in the western side of 
the SNR, as well as in the east. The observed soft X-ray flux is 
increasing more rapidly than ever, and the latest flux is four times 
brighter than three years ago. 
\end{abstract}

\section{Introduction}

We continue our monitoring program of SNR 1987A with the Advanced
CCD Imaging Spectrometer (ACIS) on board the {\it Chandra X-Ray 
Observatory}. As of 2002 December, we have performed a total of
seven observations (Table 1). Results from the first six observations 
have been reported in the literature (Burrows et al. 2000; Park et al. 
2002, 2003; Michael et al. 2002). The X-ray morphology was ring-like
and asymmetric (brighter in the east) with the emergence of 
X-ray-bright spots. The X-ray spectrum was described with a 
plane-parallel shock in a non-equilibrium ionization (NEI) state with 
an electron temperature of $kT$ $\sim$ 2.5 keV. A blast wave shock 
velocity of $v$ $\sim$ 3500 km s$^{-1}$ was derived from the detected 
X-ray line profiles, which also provided direct evidence of an 
electron-ion non-equilibrium behind the shock. The X-ray flux was 
non-linearly increasing as the blast wave approaches the dense 
inner ring. We here report on the results from the lastest {\it Chandra} 
observations of SNR 1987A. 
\begin{table}
\caption {{\it Chandra}/ACIS Observations of SNR 1987A}
\begin{tabular}{cccc}
\tableline
ObsID & Date (Age$^{a}$) & Exposure (ks) & Source counts \\
\tableline
124+1387$^{b}$ & 1999-10-06 (4609) & 116 & 690 \\
122 & 2000-01-17 (4711) & 9 & 607 \\
1967 & 2000-12-07 (5038) & 99 & 9031 \\
1044 & 2001-04-25 (5176) & 18 & 1800 \\
2831 & 2001-12-12 (5407) & 49 & 6226 \\ 
2832 & 2002-05-15 (5561) & 44 & 6429 \\
3829 & 2002-12-31 (5791) & 49 & 9274 \\ 
\tableline
\tableline
\end{tabular}\\ \\
$^{a}$~Day since the SN explosion.\\
$^{b}$~The first observation was split into two sequences, which
were combined in the analysis. The ACIS-S3+HETG was used. The ACIS-S3 
was chosen for the other observations.
\end{table}

\section{Morphology}
\begin{figure}
\centerline{\psfig{file=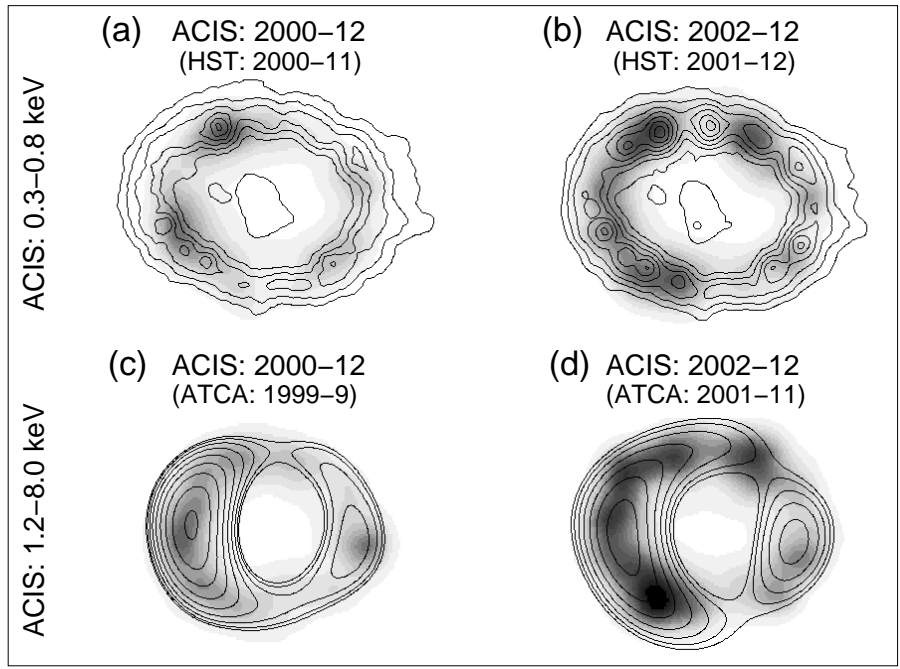,width=\textwidth}}
\caption{The exposure-corrected ACIS images of SNR 1987A. Darker 
gray-scales are higher intensities. The image deconvolution and 
smoothing, after performing a sub-pixel resolution, have been applied.  
Top panels (a, b) are the 0.3$-$0.8 keV, and bottom panels (c, d)
are the 1.2$-$8.0 keV band images. The optical and the radio contours 
are overlaid.}
\end{figure}
The ACIS images (Fig. 1) have been generated by the data reduction
process described in Burrows et al. (2000) and Park et al. (2002). 
Fig. 1 shows the overall brightening and development of new X-ray 
bright spots, particularly in the western side of the SNR. SNR 1987A 
is expected to be a complete ring in X-rays. In 2000, the strong 
correlation between the soft X-ray and the optical images was 
interpreted as emission from the slow shock encountering protrusions 
of the dense inner ring. The correlation between the hard X-ray and 
the radio images was consistent with emission from the fast shock 
propagating into an H{\small II} region interior to the ring. As of 
2002-12, correlations 
between the X-ray and the optical/radio images are more complex than 
such a simple picture, which is perhaps expected as the blast wave 
is reaching the main body of the inner ring.

\section{Spectrum and Flux}
The X-ray spectrum from each observation can be fitted with a single 
NEI plane shock model with an electron temperature of $kT$$\sim$2$-$3 
keV and low abundances (0.1$-$0.5 solar). 
Based on the data with the most significant photon statistics, we 
find that the electron temperature has decreased slightly, while the 
volume emission measure (EM) has substantially increased for the last 
two years (Fig. 2). These results are consistent with our physical 
picture of the blast wave encountering a dense circumstellar medium 
(CSM). A growing contribution to the observed spectrum from the slow 
shock is then expected (Michael et al. 2002). 
\begin{figure}
\centerline{\psfig{file=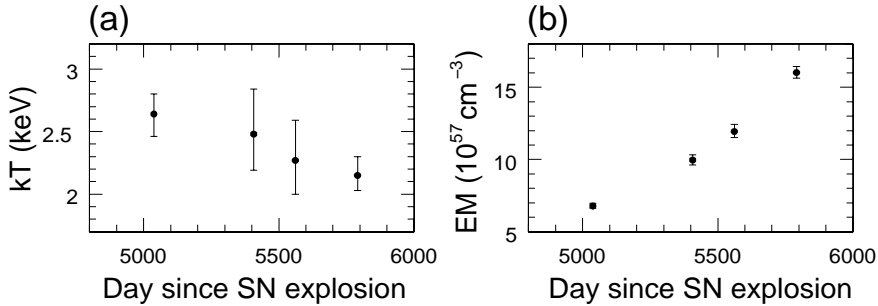,width=0.90\textwidth}}
\caption{The electron temperature and the emission measure
variations of SNR 1987A between 2000-12 and 2002-12.}
\end{figure}
A two-temperature model indeed statistically improves the fit,
particularly with the latest data (Fig. 3). The soft component 
($kT$$\sim$0.25 keV) implies a slow shock velocity ($v$$\sim$400 
km s$^{-1}$) with a highly advanced ionization state 
($n_et$$\sim$10$^{13}$ cm$^{-3}$ s). This is consistent with the 
decelerated shock front in the dense CSM. The 0.5$-$5 keV band 
flux from the soft component has tripled since 2000-12, while the 
flux from the hard component has doubled. We note that the flux 
increase from the slow shock in the western half (by a factor of
6) is much larger than in the east ($\sim$30\%). The flux increase 
from the fast shock is constant between the east and the west. 
These results suggest that the blast wave has now begun to encounter 
the dense CSM in the western side of the SNR a few years after 
it did in the eastern side.
%\section{Lightcurve}
The X-ray lightcurve shows a non-linear flux increase rate (Fig. 4). 
We have thus attempted non-linear fits to the data: e.g., in Fig. 4, 
we present a fit by assuming an exponential density distribution 
along the radius of the inner ring. The best-fit model implies an 
$\sim$20 times higher density in the inner ring than in the H{\small II} 
region.  The actual density contrast, which might be higher than this, 
could be measured as the shock-CSM interaction proceeds.

\begin{figure}
\centerline{\psfig{file=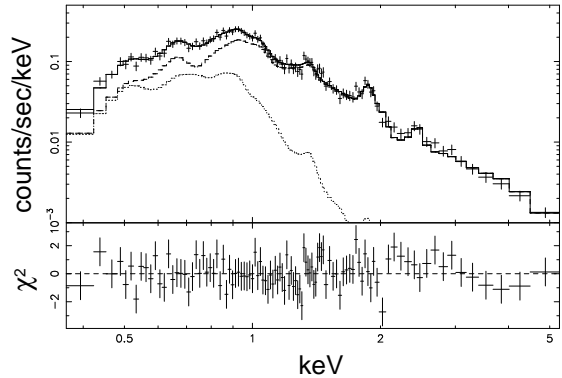,width=0.60\textwidth}}
\caption{The X-ray spectrum of SNR 1987A as of 2002-12.
The best-fit two-component plane shock model is overlaid.}
\end{figure}
%\vspace{1cm}
\begin{figure}
\centerline{\psfig{file=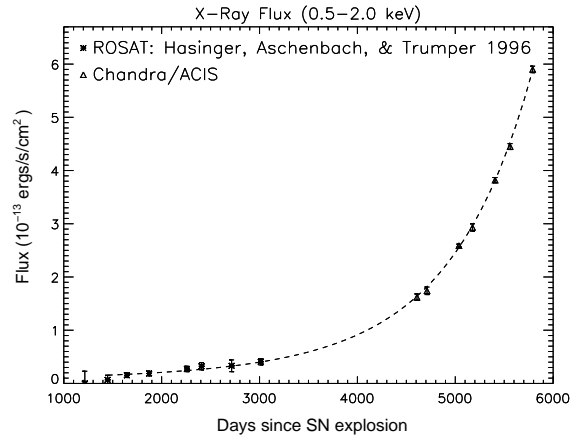,width=0.65\textwidth}}
\caption{The X-ray lightcurve of SNR 1987A. The dashed curve is a 
best-fit model assuming an exponential distribution of the radial density 
profile interior to the inner ring.}
\end{figure}
\end{document}